# A model of the 3-μm hydration band with Exponentially Modified Gaussian (EMG) profiles: application to hydrated chondrites and asteroids

S. Potin[1], S. Manigand[2], P.Beck[1,3], C. Wolters[1], B. Schmitt[1]
[1]Université Grenoble Alpes, CNRS, IPAG (414 rue de la Piscine, 38400 Saint-Martin d'Hères, France), [2]Niels Bohr Institute & Centre for Star and Planet Formation, University of Copenhagen, (Øster Voldgade 5–7, DK-1350 Copenhagen K., Denmark), [3]Institut Universitaire de France, Paris, France

**ABSTRACT**
We present here a new method to model the shape of the 3-μm absorption band in the reflectance spectra of meteorites and small bodies. The band is decomposed into several OH/$H_2O$ components using Exponentially Modified Gaussian (EMG) profiles, as well as possible organic components using Gaussian profiles when present. We compare this model to polynomial and multiple Gaussian profile fits and show that the EMGs model returns the best rendering of the shape of the band, with significantly lower residuals. We also propose as an example an algorithm to estimate the error on the band parameters using a bootstrap method. We then present an application of the model to two spectral analyses of smectites subjected to different $H_2O$ vapor pressures, and present the variations of the components with decreasing humidity. This example emphasizes the ability of this model to coherently retrieve weak bands that are hidden within much stronger ones.

### 1) Introduction

Spectroscopic evidences of hydration are common among asteroids and meteorites. The most direct evidence is the presence of a strong absorption feature detected around 2.7-3.0 μm, corresponding to vibration modes of –OH groups in hydrated minerals and $H_2O$ molecules on the surface of the small bodies (Rivkin, 2003; Usui et al., 2019). The shape of this feature is sometimes used as a classification tool for asteroid observations (Takir and Emery, 2012; Usui et al., 2019).

Several methods are already used to retrieve the shape of this hydration band. The absorption feature can be modeled with radiative transfer models using laboratory reflectance spectra of pure phases. In that case, the proportion of each is set as free parameter as well as the grain size, and sometimes its distribution (De Sanctis et al., 2015). Without using pure analogues spectra, a high-order polynomial fit can return the positions of existing components (Rivkin et al., 2019) as was used for the position of the hydrated silicates band (Fornasier et al., 2014).

We present here a new model to reconstruct the shape of the 3μm band that is motivated by the desire to better characterize the different contributors to this complex feature and understand the nature of hydration among the small bodies population.

### 2) Presentation of the model
#### a. The Exponentially Modified Gaussian profile

The Exponentially Modified Gaussian (EMG) profile, presented by Grushka (1972) is currently used to model asymmetrical chromatography peaks and is described by the following equation:

$$\text{EMG}(\lambda, h, \mu, \sigma, \tau) = \frac{h\sigma}{\tau}\sqrt{\frac{\pi}{2}} \exp\left[\frac{1}{2}\left(\frac{\sigma}{\tau}\right)^2 - \frac{\lambda - \mu}{\tau}\right] \text{erfc}\left[\frac{1}{\sqrt{2}}\left(\frac{\sigma}{\tau} - \frac{\lambda - \mu}{\sigma}\right)\right]$$

with $\lambda$ the wavelength, h, $\mu$ and $\sigma^2$ the amplitude, mean and variance of the Gaussian respectively, $\tau$ the exponent relaxation factor, and erfc(x) the complementary error function :

$$\text{erfc}(x) = \frac{2}{\sqrt{\pi}} \int_x^\infty e^{-t^2} dt$$

The EMG profile is a convolution of a Gaussian profile and an exponential decay, thus the amplitude, peak position, width and skewness of the resulting shape are not directly given by the values of h, $\mu$, $\sigma$ and $\tau$ injected in the model. *Figure 1* presents the effect of all parameters of the EMG profile on the resulting shape.

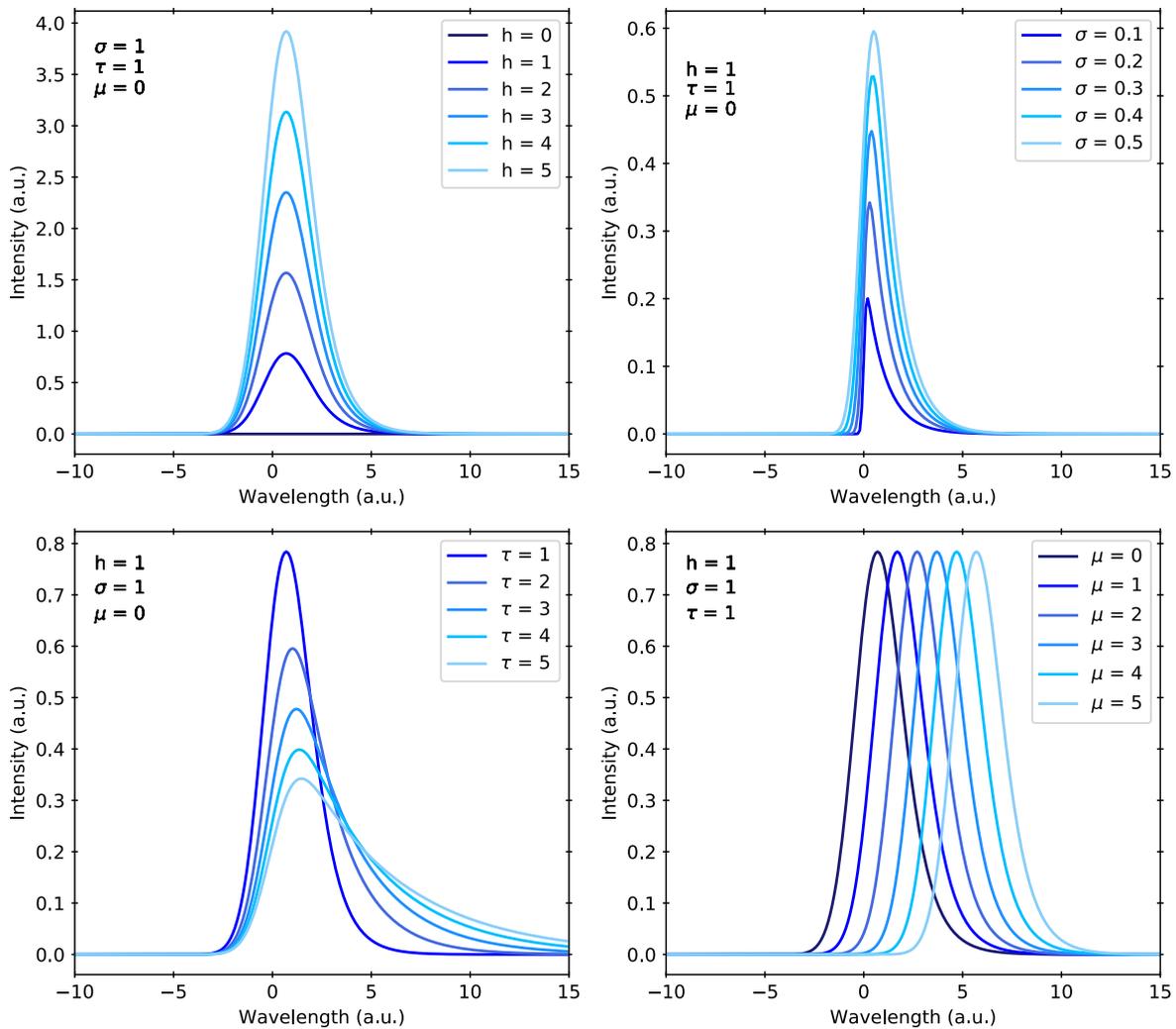

*Figure 1: Variations of the band shape resulting from the EMG model with each individual parameter. For each panel, the fixed parameters are set to 1 for h, $\sigma$ and $\tau$ and to 0 for $\mu$.*

The resulting shape is a combination of all parameters, as it is easily seen on *Figure 1*. For example, the amplitude of the component depends on the parameters h, $\sigma$ and $\tau$, and the peak position is dependent of $\tau$ and $\mu$. It is also important to note that even with $\mu$ set to 0, the position of the maximum of the peak is not at 0. This function become more symmetrical, close to a Gaussian profile when the ratio τ/σ tends to 0 (Gladney et al., 1969; Grushka, 1972).

### b. Shape of the –OH band

In order to obtain the rendering of the complete band shape, we must take into account the different carriers of the –OH groups. Bishop et al. (1994), Frost et al. (2000) and Kuligiewicz et al. (2015) highlighted the presence of 3 different types of fundamental vibration bands composing the 3-µm feature:

- vibrations of hydroxyl groups in hydrated minerals around 2.7-2.8µm, with a tenuous contribution of weakly bound interlayer water molecules weakly bound to $SiO_4$ (Kuligiewicz et al., 2015)

- symmetric and asymmetric stretching vibrations of the adsorbed (weakly bound) $H_2O$ molecules around 2.9µm

- stretching vibrations of the structural $H_2O$ molecules bound into interlayer cations (strongly bound) around 3.1µm

These components trace the presence of water and hydrated minerals in the sample, but the model must of course be adapted to the nature of each analyzed sample. For example, another component can be added to the model in case of the presence of a shoulder on the left wing of the 3-µm band (Bishop et al., 1994; Takir et al., 2013).

The exact position of each component depends on the chemical composition of the carrier (Bishop et al., 1994 and references therein). The number of components injected in the fit model must be chosen according to the nature of the sample, (e.g. dehydrated or rich in adsorbed water), and given which fit option returns the lowest residuals.

We present hereafter the modeling of several typical spectra, from both laboratory measurements and ground-based observations. The reflectance spectra of the meteorite Tagish Lake, before and after heating at 250°C, were acquired at IPAG with the spectro-gonio radiometer SHADOWS (Potin et al., 2018), while telescopic observations of the asteroids (2) Pallas and (121) Hermione were taken from the AKARI space telescope data collection (Usui et al., 2019). The spectra are presented in *Figure 2*.

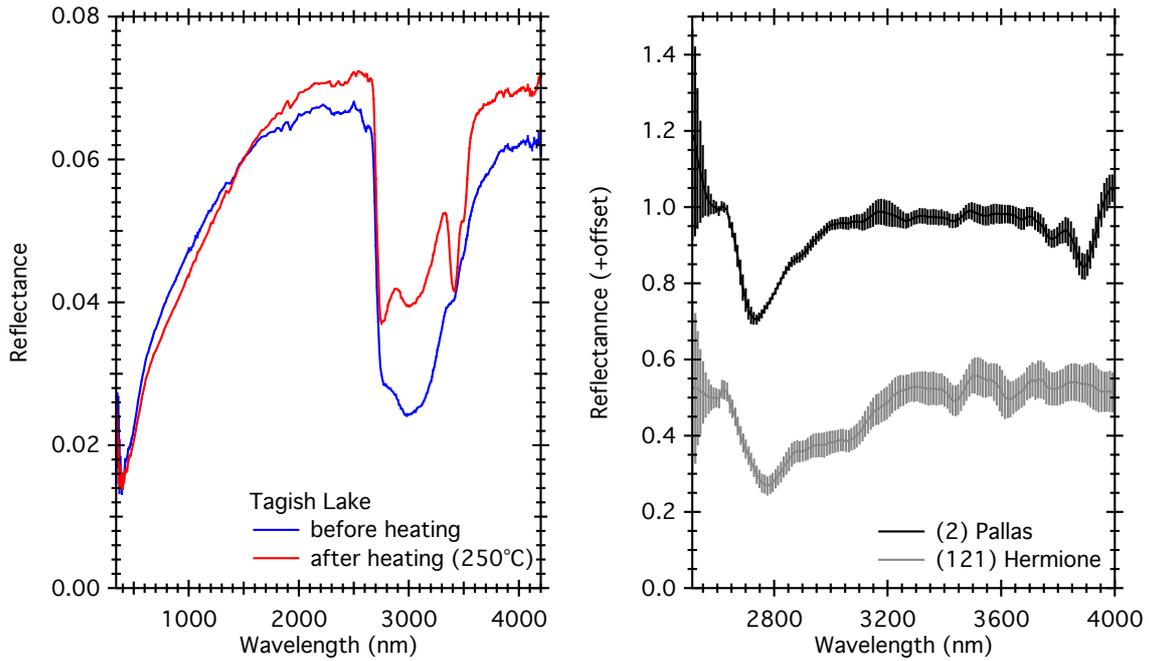

*Figure 2: Left: reflectance spectra of the Tagish Lake meteorite before (blue) and after (red) heating at 250°C. Right: reflectance spectra of the asteroids (2) Pallas and (121) Hermione. A negative offset of 0.5 has been applied to the spectrum of (121) Hermione for clarity. Shaded area correspond to spectral ranges with high calibration errors, as described in (Usui et al., 2019).*

For each spectrum, the reflectance has been normalized by the continuum, considering a linear continuum between the two inflexion points on both wings of the band. Obviously, the shape given to the continuum must be adapted to each case. A linear continuum is appropriate for relatively dry samples where the reflectance returns to the continuum on each side of the band. In case of humid samples, or complex samples with a major contribution of organics or carbonates around 4 µm, the 3-µm band will present a shoulder on the low-wavelength wing, and the reflectance will not reach the continuum after the low-wavelength wing. In these cases, the use of a linear continuum is not appropriate, and will change the shape the band after its removal. A linear continuum could be used in each of our spectra presented here.

Absorption bands of –$CH_2$ and –$CH_3$ groups in carbonaceous materials can be detected on the reflectance spectra around 3.4 µm and thus on the right wing of the 3-µm band. Choice can be made of simply removing these bands from the data, but as an example we decided to take their absorptions into account. Note that the feature of the organic compounds is generally composed of 4 or 5 absorption bands (Orthous-Daunay et al., 2013), but in the case of Tagish Lake, the spectral resolution of the goniometer does not allow the distinction of all components and thus we assimilated them as two Gaussians profiles. The complete model of the band contains one EMG model for each components of the 3-µm band, and Gaussian profiles for the organics absorption bands:

$$\text{FIT}(\lambda) = 1 - \left[\sum_{i=1}^{i=N} \text{EMG}(\lambda, h_i, \mu_i, \sigma_i, \tau_i) + \text{Gaussian}(\lambda, A_{\text{orga 1}}, \sigma_{\text{orga 1}}, \lambda_{0\ \text{orga1}}) \right.$$
$$\left. + \text{Gaussian}(\lambda, A_{\text{orga 2}}, \sigma_{\text{orga 2}}, \lambda_{0\ \text{orga2}}) \right]$$

with N number of the components carrying the –OH groups, and with:

$$\text{Gaussian}(\lambda, A_{\text{orga}}, \sigma_{\text{orga}}, \lambda_{0\ \text{orga}}) = A_{\text{orga}}\ \frac{1}{\sigma_{\text{orga}}\sqrt{2\pi}}\ \exp\left[-\frac{(x - \lambda_{0\ \text{orga}})^2}{2\sigma_{\text{orga}}^2}\right]$$

with $A_{\text{orga}}$, $\sigma_{\text{orga}}$ and $\lambda_{0\ \text{orga}}$ the amplitude, broadness and center of the Gaussian profile for the organics respectively. If the reflectance spectrum is not normalized, the 1 at the beginning of the model (previous equation) can be changed to a linear continuum $A\lambda + B$ with A the value of the spectral slope and B the reflectance of the continuum.

For each point, the residuals are calculated as:

$$\text{Residuals}(\lambda) = \text{Refl}(\lambda) - \text{FIT}(\lambda)$$

with FIT($\lambda$) the modeled reflectance calculated using the multiple EMGs model and Refl($\lambda$) the measured reflectance at the wavelength $\lambda$.

*Figure 3* presents the best fit models of the reflectance spectra of the meteorite Tagish Lake before and after heating at 250°C and the asteroids (2) Pallas and (121) Hermione. For each fit, the residuals, difference between the observed reflectance and the model, are plotted underneath.

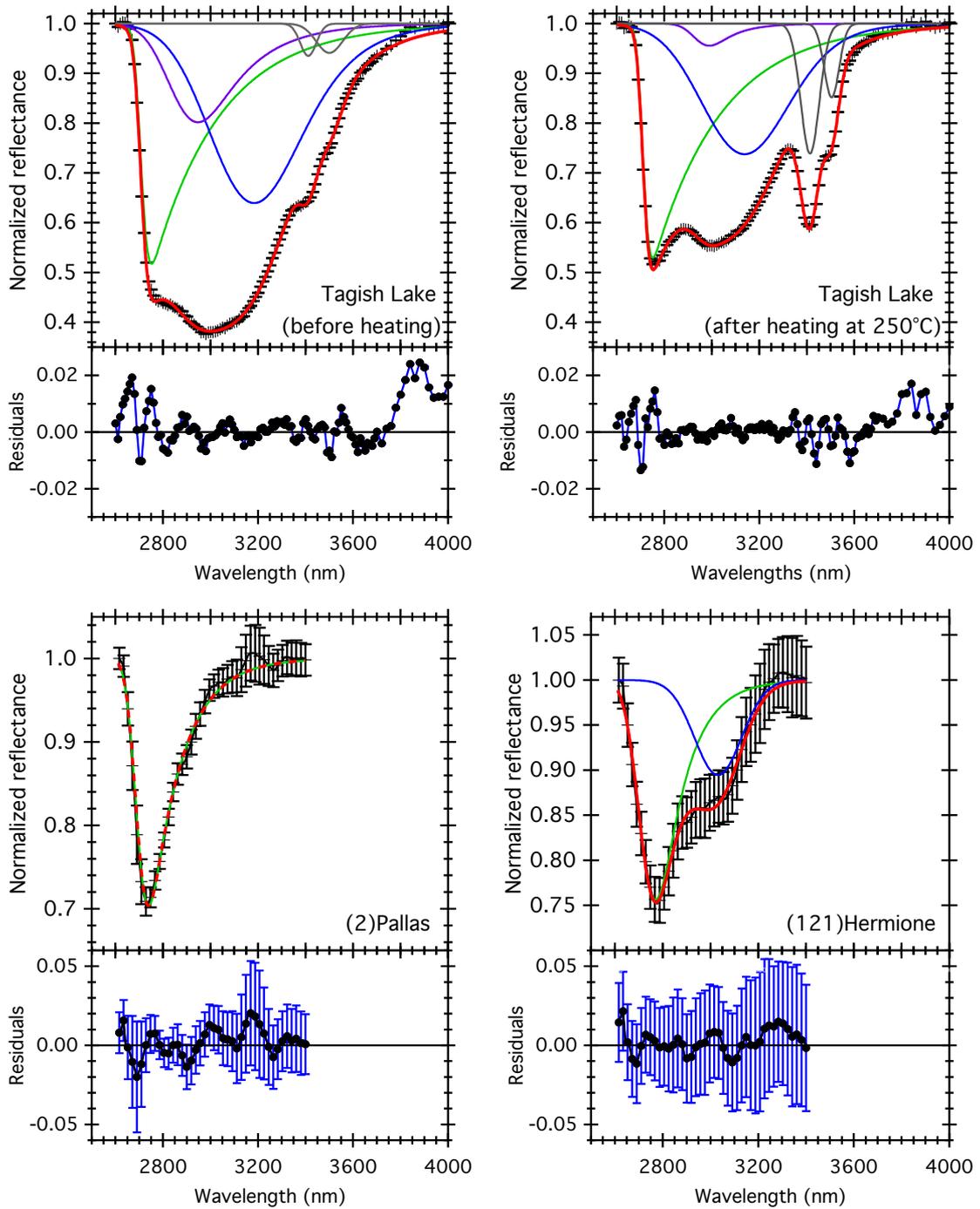

*Figure 3: Modeled reflectance spectra of Tagish Lake before heating (top left), after heating at 250°C (top right), and the asteroids (2) Pallas (bottom left) and (121) Hermione (bottom right). Black crosses with error bars: reflectance data. Red: resulting multiple EMGs model fit. Green: -OH stretching mode in hydrated minerals. Purple: $H_2O$ stretching modes in weakly bound adsorbed water. Blue: $H_2O$ stretching modes in strongly bound water molecules. Grey: organic features. Note that the fit of the reflectance spectrum of (2) Pallas uses only the hydrated minerals component.*

The parameters h, σ, τ and μ of the modeled EMGs profiles are presented in *Table 1*. The corresponding band parameters of the components are also given.

*Table 1: Parameters of the modeled EMGs profiles derived from the model and corresponding band parameters.*

|  |  |  | Tagish Lake (unheated) | Tagish Lake (heated) | Pallas | Hermione |
|---|---|---|---|---|---|---|
| First component | Fit values | h | 2.823 | 3.096 | 0.694 | 0.404 |
|  |  | σ | 24.283 | 21.186 | 36.491 | 54.220 |
|  |  | τ | 292.859 | 299.631 | 131.528 | 106.456 |
|  |  | μ | 2706.502 | 2709.251 | 2691.400 | 2714.935 |
|  | Derived components parameters | Band depth (%) | 48.5 | 46.5 | 29.6 | 24.5 |
|  |  | Position (nm) | 2752.1 | 2750.5 | 2739.5 | 2769.4 |
|  |  | FWHM (nm) | 263.1 | 261.0 | 172.7 | 202.9 |
| Second component | Fit values | h | 0.464 | 0.318 | - | 0.107 |
|  |  | σ | 154.565 | 160.991 | - | 98.680 |
|  |  | τ | 169.074 | 123.288 | - | 2.927 |
|  |  | μ | 3072.097 | 3034.698 | - | 3025.928 |
|  | Derived components parameters | Band depth (%) | 35.4 | 26.9 | - | 9.9 |
|  |  | Position (nm) | 3185.9 | 3129.0 | - | 3034.9 |
|  |  | FWHM (nm) | 456.8 | 439.8 | - | 216.2 |
| Third component | Fit values | h | 0.347 | 0.0569 | - |  |
|  |  | σ | 94.589 | 33.898 | - | - |
|  |  | τ | 185.225 | 74.129 | - | - |
|  |  | μ | 2851.987 | 2942.692 | - | - |
|  | Derived components parameters | Band depth (%) | 20.9 | 3.2 | - | - |
|  |  | Position (nm) | 2947.3 | 2978.6 | - | - |
|  |  | FWHM (nm) | 336.5 | 126.0 | - | - |

The multiple EMGs model results in a good description of the whole 3-µm band shape of each studied spectra, with only one, two or three components. For all these fits, the residuals are lower than ±0.02 of the original measurement, and centered around 0. For Tagish Lake, the deviation of the fit in the outer wing at wavelengths higher than 3700 nm can be due to the continuum normalization and does not affect the best-fit solution of the complete band. It is important to note that this model can also be used for transmission spectroscopy, switching from absorption feature to absorbance peak.

To compare the multiple EMGs model with existing methods, *Figure 4* presents the spectral modeling of the meteorite Tagish Lake spectrum, using a polynomial profile and a sum of Gaussian profiles.

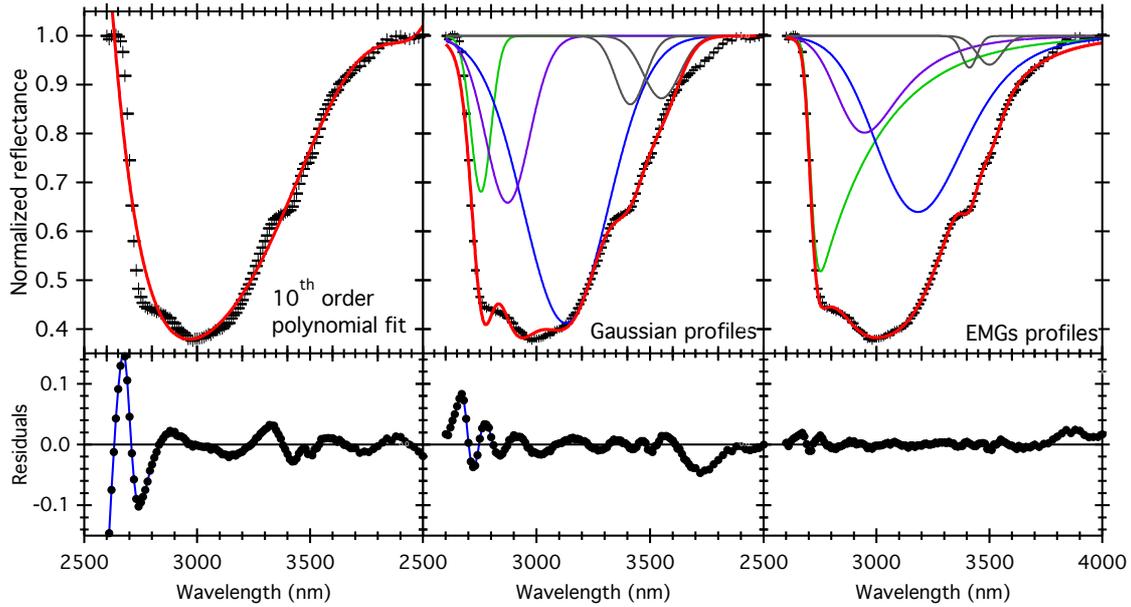

*Figure 4: Spectral modeling of the 3µm band of the meteorite Tagish Lake (unheated) with a 10th order polynomial fit (left), a sum of Gaussian profiles (center), and with the EMGs profiles (right). Representation are the same as Figure 3.*

The derived band parameters of each component are presented in *Table 2*.

*Table 2: Band parameters of the components derived from the models using Gaussians or EMGs profiles.*

|  |  | Gaussians | EMGs |
|---|---|---|---|
| First component | Band depth (%) | 31.913 ± 0.011 | 48.556 ± 0.017 |
|  | Position (nm) | 2755.353 ± 0.017 | 2752.064 ± 0.027 |
|  | FWHM (nm) | 104.4961 ± 0.046 | 263.11 ± 0.13 |
| Second component | Band depth (%) | 58.758 ± 0.006 | 35.386 ± 0.018 |
|  | Position (nm) | 3127.971 ± 0.052 | 3185.96 ± 0.11 |
|  | FWHM (nm) | 435.75 ± 0.10 | 456.77 ± 0.26 |
| Third component | Band depth (%) | 34.127 ± 0.010 | 20.989 ± 0.026 |
|  | Position (nm) | 2872.211 ± 0.050 | 2947.31 ± 0.12 |
|  | FWHM (nm) | 228.052 ± 0.046 | 336.47 ± 0.15 |

The high order polynomial model can reproduce the shape of the band but only over a limited spectral range. Though the right wing and the bottom of the band are rather well defined, a tenth order polynomial fit cannot model the whole feature satisfyingly. In addition, the resulting parameters of the polynomial model have no physical interpretation. On the other hand, both Gaussians and EMGs profiles return band parameters for the components, but their values are different and the representation of the three OH/$H_2O$ components as Gaussian profiles does not accurately reproduce the steep left wing and the bottom of the band. More than three Gaussian components would be necessary to model the whole feature, which tends to disagree with the spectroscopic interpretation of the band.

If the 3-µm band is wide enough to include the organics features, as it is the case for Tagish Lake spectra, these absorption bands will contribute to the shape of the complete band. *Figure 5* presents the modeled spectrum of Tagish Lake (unheated) with and without taking into account the organic features.

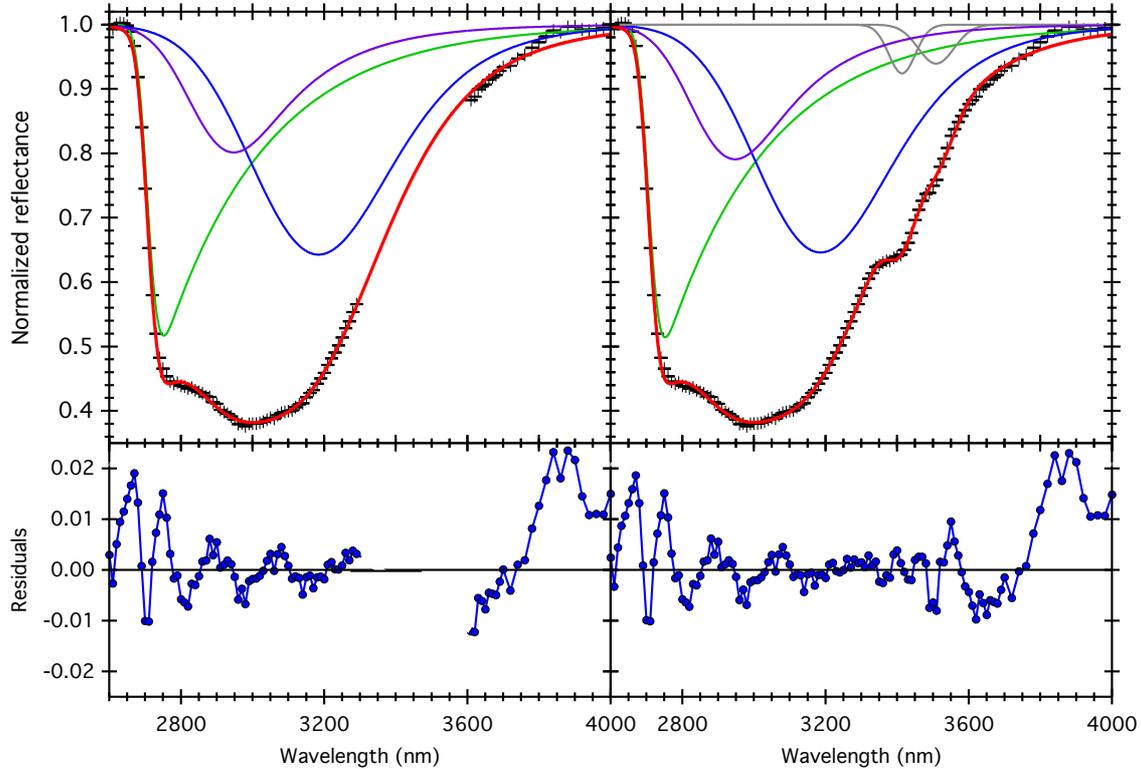

*Figure 5: Modeled spectrum and corresponding residuals of the meteorite Tagish Lake (unheated). Right: after removal of the spectral range of the organic features, Left: taking into account the organic features.*

The derived parameters of the components are presented in *Table 3*.

*Table 3: Band parameters of the complete absorption band of the Tagish Lake meteorite (unheated) and the different components derived from the modeled spectrum with and without taking into account the organic features.*

|  |  | Complete band | First component | Second component | Third component | Organic 1 | Organic 2 |
|---|---|---|---|---|---|---|---|
| Without organics | Band depth (%) | 61.8029 ± 0.0032 | 48.301 ± 0.015 | 35.716 ± 0.018 | 13.838 ± 0.016 | - | - |
| | Position (nm) | 3000.877 ± 0.085 | 2751.910 ± 0.028 | 3183.872 ± 0.088 | 2947.18 ± 0.10 | - | - |
| | FWHM (nm) | 679.54 ± 0.060 | 261.66 ± 0.14 | 470.36 ± 0.21 | 340.79 ± 0.12 | - | - |
| With organics | Band depth (%) | 61.7749 ± 0.0031 | 48.556 ± 0.017 | 35.386 ± 0.018 | 20.898 ± 0.026 | 7.562 ± 0.011 | 6.042 ± 0.011 |
| | Position (nm) | 3000.13 ± 0.10 | 2752.064 ± 0.027 | 3185.96 ± 0.11 | 2947.31 ± 0.12 | 3413.606 ± 0.081 | 3508.90 ± 0.15 |
| | FWHM (nm) | 740.98 ± 0.045 | 263.11 ± 0.13 | 456.77 ± 0.26 | 336.47 ± 0.15 | 87.58 ± 0.19 | 117.74 ± 0.21 |

The removal of the organic features in the model does not induce significant changes in the derived components, only the FWHM of the second component and the amplitude of the third are impacted. The FWHM of the second component is reduced from 470 nm to 457 nm, and the amplitude of the third component increases from 14 % to 21 %.

Removing a range of the spectrum results in a less constrained model, thus a relatively different rendering of the shape. If possible, all features impacting the 3-µm band must be modeled to ensure a proper rendering of the shape.

### c.     Band parameters and errors calculations

Asteroids are classified according to the shape of their reflectance spectra, and astronomical observations permit the precise determination of the position of the minimum of the 3-µm band, its amplitude and sometimes its shape (Ammannito et al., 2016; Beck et al., 2010; Berg et al., 2016). Spectral modeling with the multiple EMGs model enables a precise determination of each band parameters, namely the minimum position, the amplitude and the FWHM, and their respective errors. As the shape of the whole 3-µm band is a combination of up to three EMGs components (plus two Gaussian profiles if the organic features are detected), their band parameters (depth, width and wavelength at minimum) can be retrieved from the EMG model parameters.

The combination of the EMG model fitting and the band parameters calculation makes the errors estimation difficult to set. Here we propose an example of algorithm using a statistical method called bootstrap. This method consists in reproducing the fit and calculating the components parameters a large number of times on the original data to whom a small random fluctuation is added, based on the error of the data. This produces a large sample of each component parameters, which distribution is centered on the most probable value, thus the best fit, with a FWHM corresponding to the error on the component parameter.

## 3)     Example of application

We applied the model to spectroscopic data published in Kuligiewicz et al., (2015). They performed infrared transmission measurements centered on the 3-µm band on smectites saturated with $H_2O$ under varying humidity (ratio of the partial vapor pressure of $H_2O$ surrounding the sample by the saturation vapor pressure at 25°C). The complete sets of spectra are presented in *Figure 6* and *Figure 7*. In order to retrieve an error on the component parameters, we fixed the error on the measured transmission to 0.001.

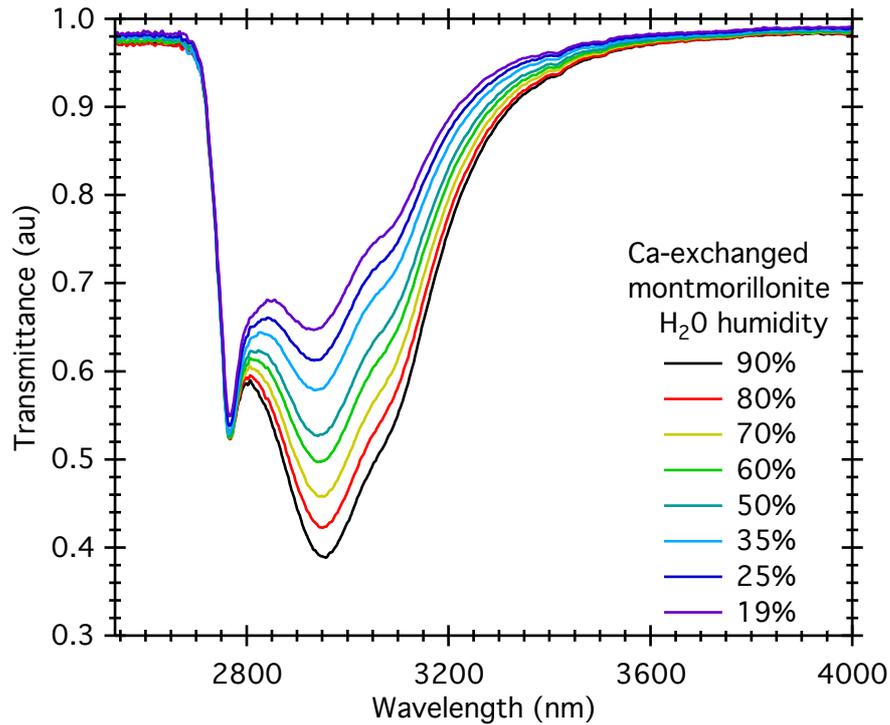

*Figure 6: Transmission spectra of the Ca-exchanged montmorillonite under varying $H_2O$ humidity levels. Data from Kuligiewicz et al. (2015), available in the LSD @ SSHADE database (Kuligiewicz et al., 2014a).*

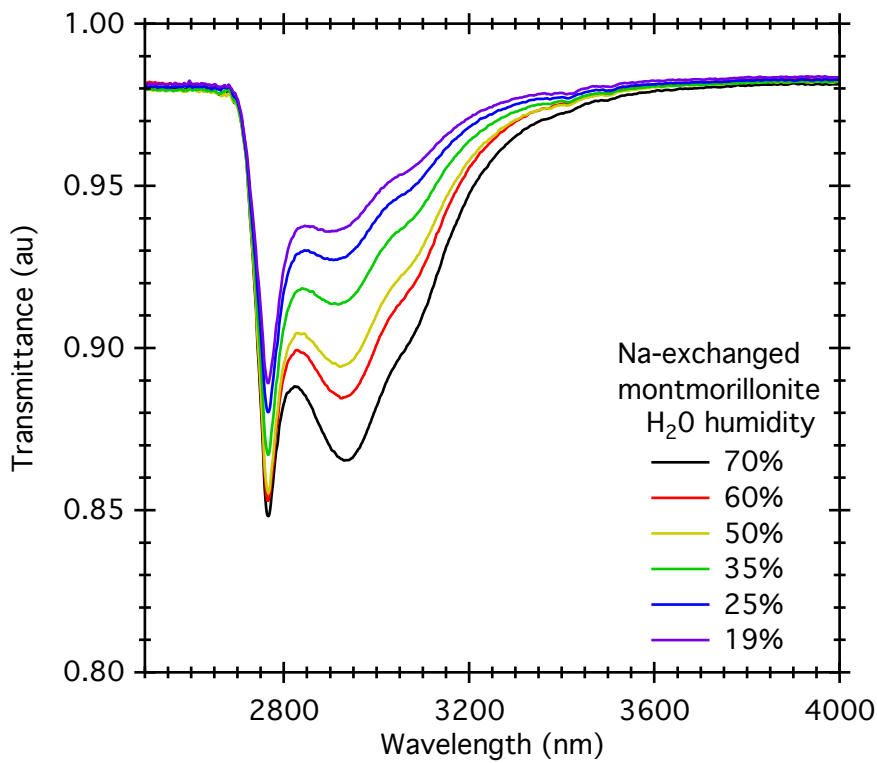

*Figure 7: Transmission spectra of the Na-exchanged montmorillonite under varying $H_2O$ humidity levels. Data from Kuligiewicz et al. (2015), available in the LSD @ SSHADE database (Kuligiewicz et al., 2014b).*

For each spectrum, we tried several numbers of components for the model. We kept the option that returned the least residuals. An example is shown in *Figure 8*.

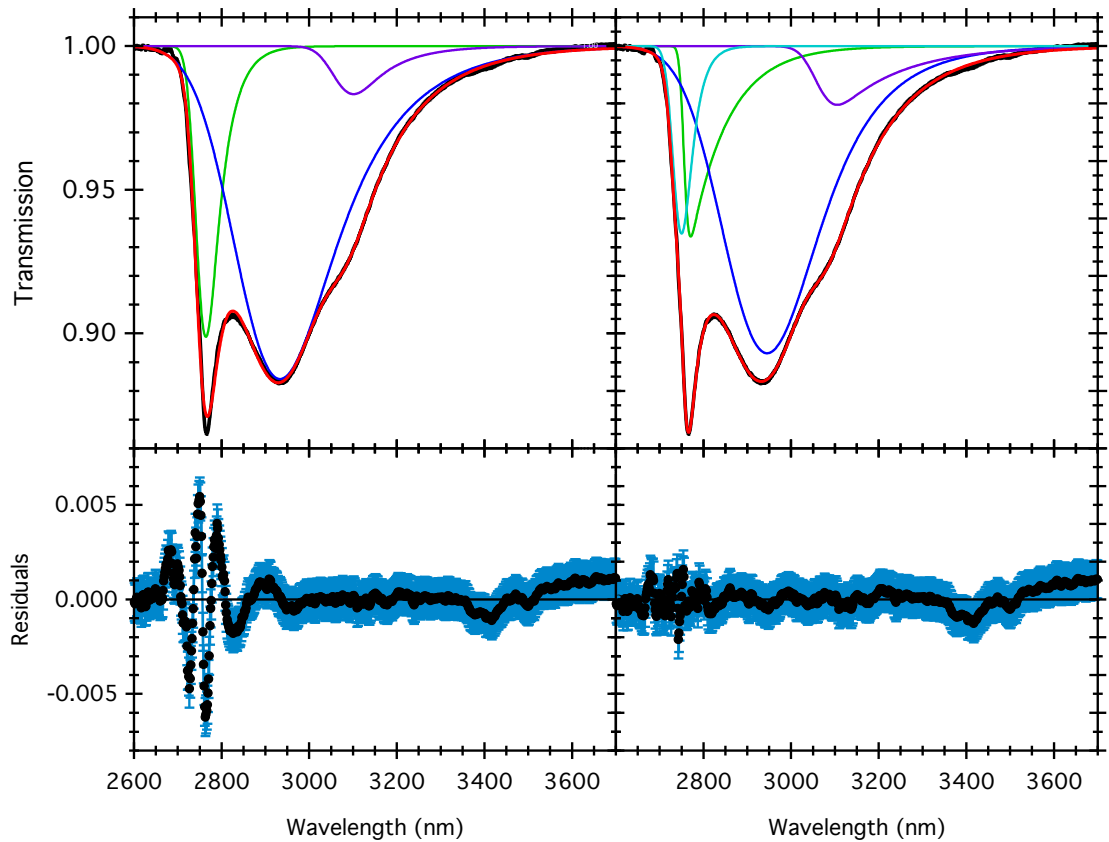

*Figure 8: Transmission spectrum of Na-exchanged montmorillonite with 70% $H_2O$ humidity, modeled with 3 (left) and 4 (right) EMGs components in the model. Black: measured transmission. Red: modeled spectrum. Green, Cyan, Blue and Purple: components of the model.*

The presence of three inflexions in the spectrum, around 2760 nm, 2940 nm and 3120 nm first led to a model using three components. However, this option returned an inaccurate fit of the 2760 nm band, with residuals peaking to more than ±0.006. Adding a fourth component to the model, representing the weakly bound interlayer water molecules close to the metal-OH component, returned a significantly better modeled spectrum. The residuals are lower than ±0.002 over the whole spectral range. For this set of spectra, the option of 4 component was selected and kept for the analysis.

*Figure 9* and *Figure 10* show the deconvolution of the 3-µm band into four components, for both Ca-exchanged and Na-exchanged montmorillonite.

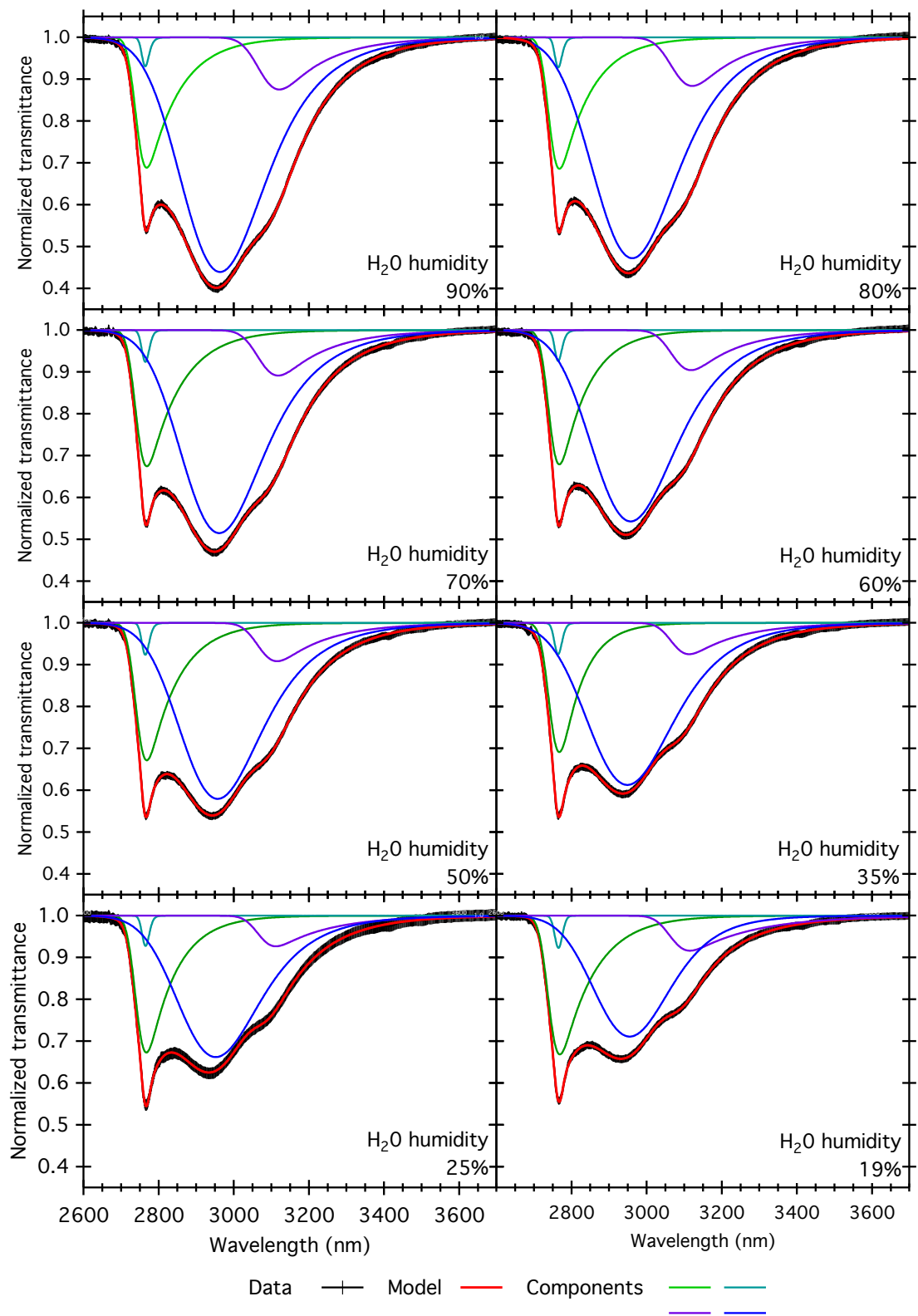

*Figure 9: Modeled spectra of Ca-exchanged montmorillonite at different humidity levels. Black: original measurement, Red: modeled spectrum, Green, Cyan, Blue and Purple: components of the model.*

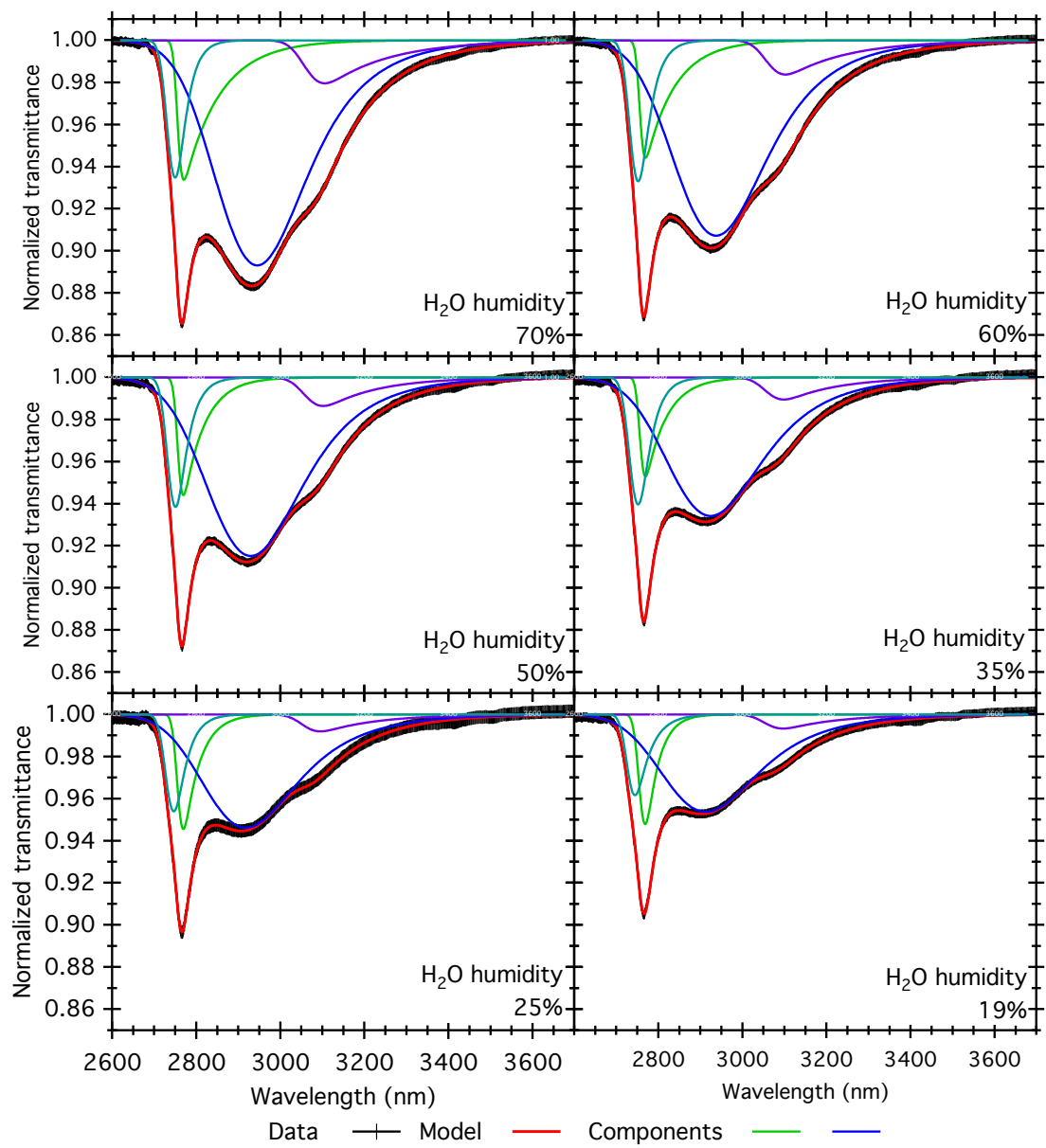

*Figure 10: Modeled spectra of Na-exchanged montmorillonite at different humidity levels. Black: original measurement, Red: modeled spectrum, Green, Cyan, Blue and Purple: components of the model.*

### a. Interaction between the components

The measured absorption band corresponds to the sum of the absorption of each components. All of them participate in the global shape of the feature. A shift is observed between the detected minima and inflexion points of the band, and the fit position of the components, as seen in *Figure 11*.

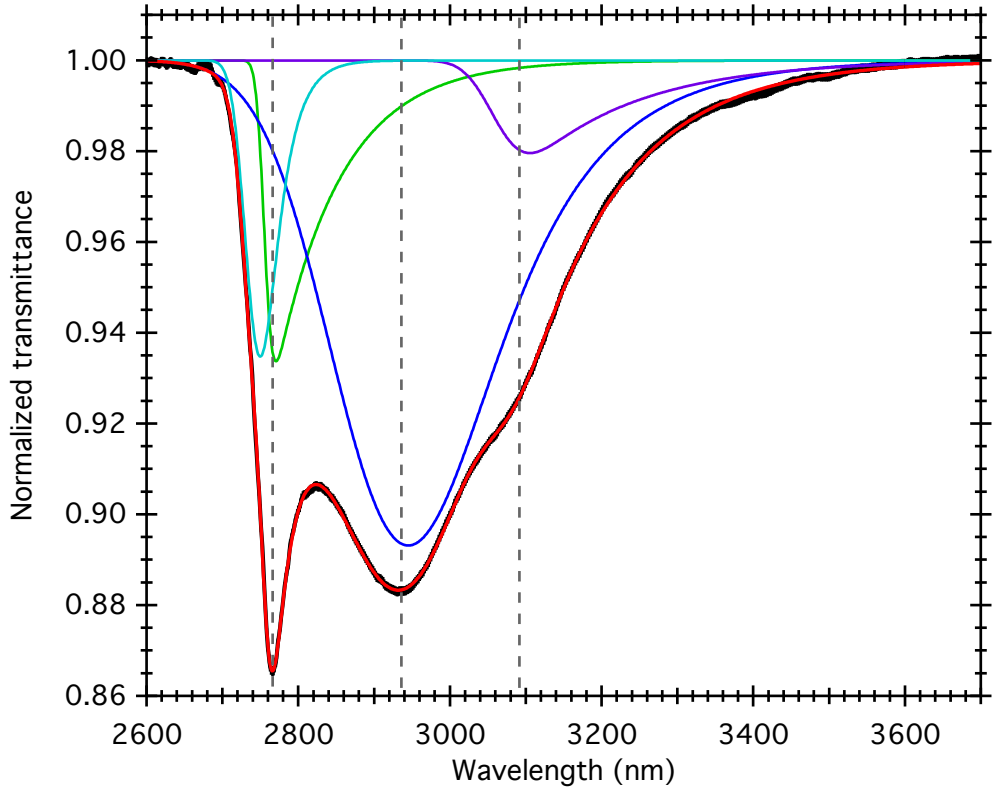

*Figure 11: Transmission spectrum of Na-exchanged montmorillonite with 70% of $H_2O$ humidity. The color code is similar to Figure 10. The grey doted lines represent the position of the detected minima and inflexion points of the global absorption band.*

It is noticeable on *Figure 11* that the positions of the detected minima and of the modeled components do not match. *Table 4* compares their positions.

*Table 4: Minima of transmission detected on the complete band (measured on the spectrum), compared to the position of the modeled components. The spectral step of the measurement is taken as the error on the position. The errors on the modeled components are computed assuming an error of 0.001 in transmission on each measurement point. The position of the third inflexion point of the global band has been determined using the derivative.*

| Detected on complete band | 2765.6 ± 1.5 | 2929.6 ± 1.7 | 3093.9 ± 1.8 |
|---|---|---|---|
| Modeled components | 2730 ± 17<br>2770 ± 1 | 2943.5 ± 2.5 | 3105.1 ± 2.2 |

These shifts can be explained by the fact that the absorption band is a convolution of several components close to one another, the minimum of a component can coincide with the wing of another. The increasing or decreasing of absorption along the wing of a component, added to the minimum of transmission of another, create an apparent minimum of transmission shifted by a few tens of nanometers from the actual component, in the direction of the other component.

### b. Evolution of the components with decreasing H$_2$O humidity

Having separated the components using the EMGs model, we can now analyze their evolution versus H$_2$O humidity. For both series, Na-exchanged and Ca-exchanged montmorillonites, the variation of the band depth and minimum position for each components are presented in *Figure 12* and *Figure 13*.

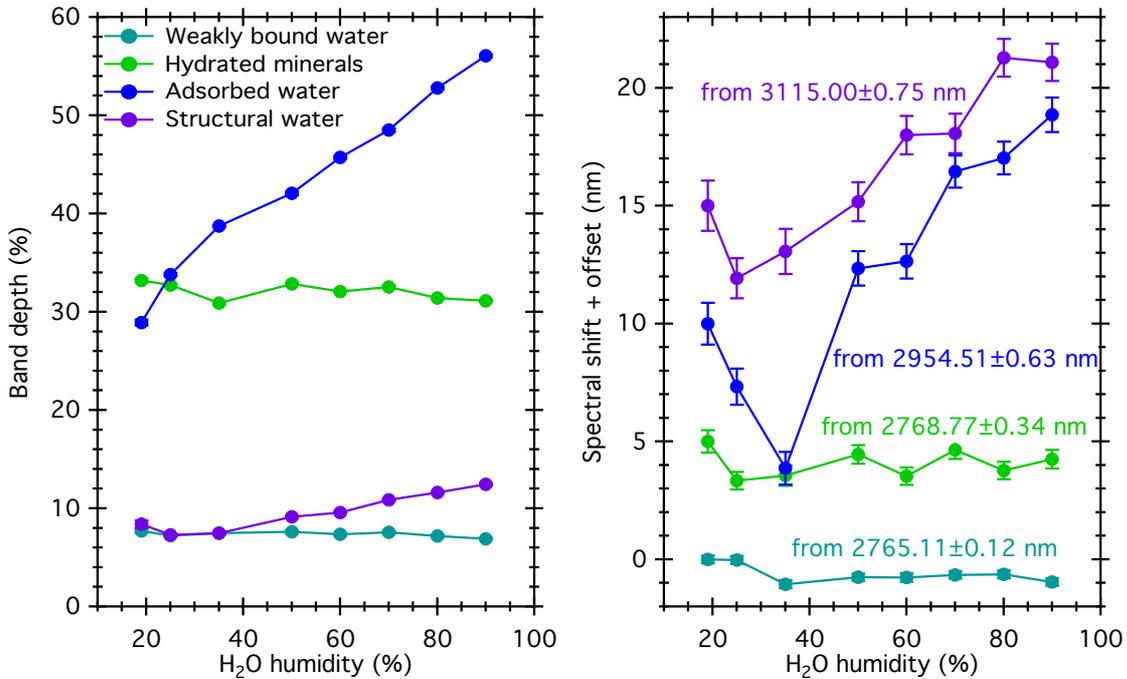

*Figure 12: Variations of the band depth and shift of the position of the minimum with varying H$_2$O humidity for each component of the 3-µm band of Ca-exchanged montmorillonite. The colors of each component are similar to Figure 9. The shift is calculated from the position derived from spectrum acquired at the lowest humidity, noted on the figure. The errorbars on the left panel are smaller than the size of the markers.*

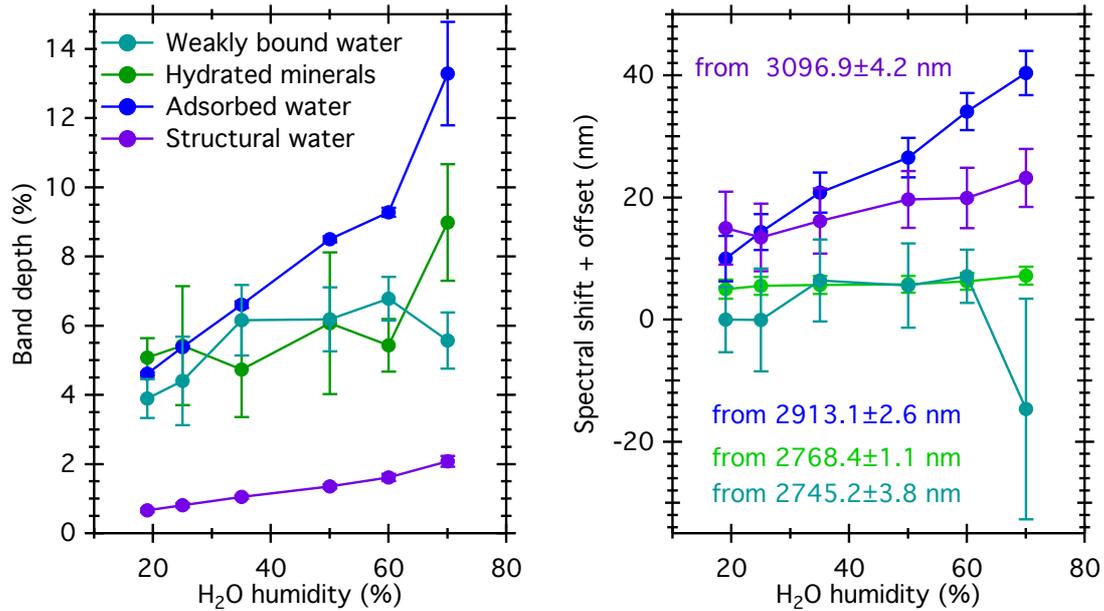

*Figure 13: Variations of the band depth and shift of the position of the minimum with varying $H_2O$ humidity for each component of the 3-µm band of Na-exchanged montmorillonite. The colors of each component are similar to Figure 10. The shift is calculated from the position derived from spectrum acquired at the lowest humidity, noted on the figure.*

On both series, we observe an increase of the band depth of the adsorbed and structural water components coupled with a shift toward longer wavelengths. This result is expected and consistent with Bishop et al. (1994), Frost et al. (2000), and Kuligiewicz et al. (2015). Schultz (1957) explained this effect as increase in H-bounding of the $H_2O$ molecules with increasing humidity, thus number of water molecules inside the sample. This induces a lowering of the vibrating frequency, and so a shift of the component toward longer wavelengths. The hydrated minerals and weakly bound water components do not show significant variations, with a maximum shift of less than 2 nm for the Ca-exchanged, and within the error bars for the Na-exchanged montmorillonite. This is exactly what is expected for the behavior of both components. In particular our deconvolution of the band, which only constrain the number of components, allows the detection of the presence of a weak component in Ca-exchanged montmorillonite with stable position and intensity whatever is the humidity. This 'hidden' component of water has been recently identified in $D_2O$ experiments as due to a population of water molecules located in the immediate vicinity of the siloxane surface and interacting very weakly with it (kuligiewicz et al. 2015). Its position for $H_2O$ was expected around 2755±12 nm, well within the range of positions, 2746 - 2764 nm, we found for this component in both montmorillonites.

### 4)    Limitations of the model

In case of ground-based observations of small bodies, the detection of the sharp low wavelength fall of the 3-µm band is impossible due to the atmospheric absorption between 2500 and 2850 nm (Rivkin et al., 2019). The left wing and the minimum of reflectance of the first component are missing from the spectrum, while its right wing still contributes to the complete band, along with the other components if any. The sharp fall represented by the left wing of the band strongly constrains the model, as it determines the asymmetric shape of the first component, as well as it minimum position. *Figure **14*** presents the modeled spectrum of (121) Hermione using the complete data, and on a partial spectrum simulating atmospheric window for ground based observations. In this last case, a first model was run with a constraint on the position of the first component: an upper limit of 2650 nm has been applied to µ in order to fix its position at a wavelength shorter than 2800 nm. A second model was then run without this constraint.

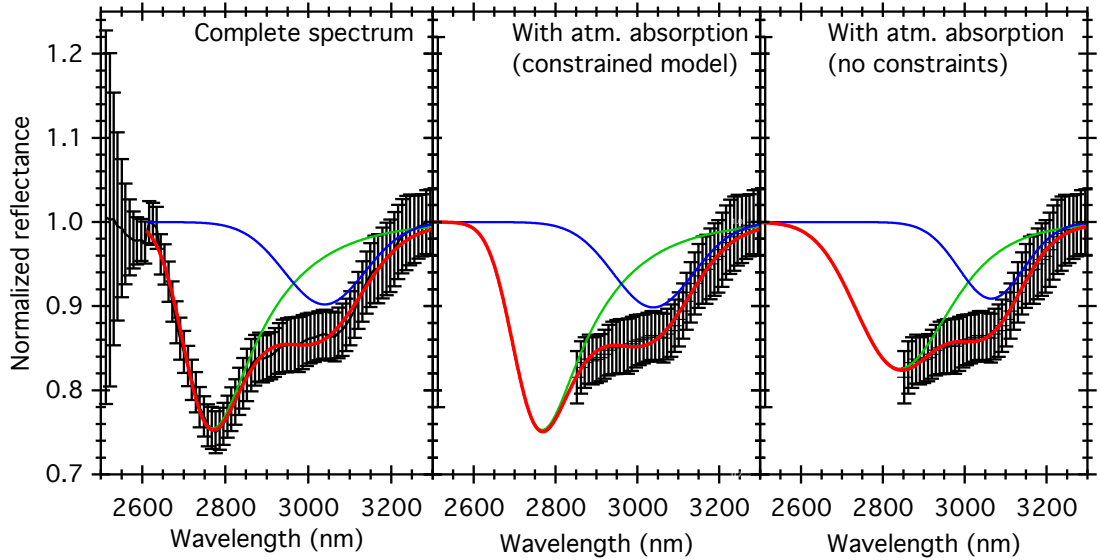

*Figure 14: Modeled reflectance of (121) Hermione with the complete spectrum (left) and with reduced data range simulating the atmospheric window (absorption between 2500 and 2850 nm), with a constraint applied on the position of the first component (middle), and without any constraint (right).*

For each model, the parameters of the components are presented in *Table **5***.

*Table 5: Amplitude and minimum position of the complete band and of the components derived from the complete spectrum and the reduced data range to simulate the atmospheric window, with and without constraints on the first component.*

| Data | | Complete band | First component | Second component |
|---|---|---|---|---|
| Full spectrum | Band depth (%) | 24.84 ± 0.99 | 24.5 ± 1.2 | 10.1 ± 2.6 |
| | Position (nm) | 2771.5 ± 7.2 | 2768.9 ± 8.3 | 3037 ± 28 |
| With atm. abs. (constrained) | Band depth (%) | 20.6 ± 9.9 | 20.1 ± 6.2 | 10.0± 3.2 |
| | Position (nm) | 2756 ± 50 | 2755 ± 54 | 3040 ± 23 |
| With atm. abs. (free) | Band depth (%) | 17.8 ± 2.5 | 17.5 ± 2.2 | 10.2 ± 4.2 |
| | Position (nm) | 2869 ± 43 | 2866 ± 41 | 3062 ± 29 |

As expected, the lack of a part of the band due to the atmospheric absorption induces significant errors on the amplitude and position of the first component, as well as on those of the complete band. Without constraints, the model does not return a component close to 2770 nm, but uses the first available data point as minimum of reflectance, and therefore fixes the position of the first component there. Constraining the model to place the first component below 2800 nm returns an apparent well-defined band, but of course its position and amplitude, and thus of complete band, bears significant errors. The error on the amplitude of the complete band is multiplied by an order of magnitude compared to the model on the full spectrum. The position of the feature is coupled with an error of 50 nm for the constrained model, but of only 7.2 nm with the full spectrum.

However, a constrained model cannot be considered anymore as representative of the composition of the studied surface. It can be only used to render the shape of the band in the spectral range lacking measurements, but is not to be considered as relevant in a scientific analysis of the composition of the studied sample or surface.

### 5)   Conclusion

Multiple EMG profiles enable a good rendering of the shape of the 3-µm band. Each EMG profile traces the presence of a different type of hydration (hydrated mineral, strongly or weakly bound water molecules) though the precise cause of the skewness is still to be explained. This model uses no other data than the spectrum to fit and can be associated with other profiles and continua on a wider spectral range. Full band parameters such as amplitude, minimum position and broadness can be easily retrieved from the modeled spectrum, taking into account the data errors. It showed its ability to coherently retrieve weak bands that are hidden within much stronger ones. This general model can be used on laboratory measurements, reflectance and transmission, as well as astronomical observations.

**Aknowledgements**

The authors would like to thank the two anonymous reviewers whose comments and suggestions significantly improved the strength of this paper.

This research is based on observations with AKARI, a JAXA project with the participation of ESA. SP is supported by the Université Grenoble Alpes (UGA) (IRS IDEX/UGA). SM is supported by the H2020 European Research Council (ERC) (grant agreement No 646908) through ERC Consolidator Grant "S4F". Research at Centre for Star and Planet Formation is funded by the Danish National Research Foundation. PB acknowledges funding from the H2020 European Research Council (ERC) (SOLARYS ERC-CoG2017_771691). CW acknowledges a PhD fellowship from CNES/ANR (ANR-16-CE29-0015 2016-2021). The instrument SHADOWS was founded by the OSUG@2020 Labex (Grant ANR10 LABX56), by 'Europlanet 2020 RI' within the European Union's Horizon 2020 research and innovation program (grant N° 654208) and by the Centre National d'Etudes Spatiales (CNES).